\documentclass[pra,aps,showpacs,floatfix,twocolumn,nofootinbib,superscriptaddress]{revtex4-1}
\usepackage{amsmath}
\usepackage{amsfonts}
\usepackage{amssymb}
\usepackage{revsymb}
\usepackage{graphicx}
\usepackage{mhchem}
\usepackage{siunitx}
\usepackage{bm}
\usepackage{dcolumn}
\usepackage{multirow}

\begin{document}
\title{Polarizability assessments of ion-based optical clocks}
\author{M. D. Barrett}
\email{phybmd@nus.edu.sg}
\affiliation{Centre for Quantum Technologies, National University of Singapore, 3 Science Drive 2, 117543 Singapore}
\affiliation{Department of Physics, National University of Singapore, 2 Science Drive 3, 117551 Singapore}
\author{K. J. Arnold}
\affiliation{Centre for Quantum Technologies, National University of Singapore, 3 Science Drive 2, 117543 Singapore}
\author{M. S. Safronova}
\affiliation{Department of Physics and Astronomy, University of Delaware, Newark, Delaware 19716, USA}
\affiliation{Joint Quantum Institute, National Institute of Standards and Technology and the University of Maryland,
College Park, Maryland, 20742}
\begin{abstract}
It is shown that the dynamic differential scalar polarisability of the $S_{1/2}-D_{5/2}$ transition in $^{138}$Ba$^+$ can be determined to an inaccuracy below $0.5\%$ across a wide wavelength range ($\lambda>700\,\mathrm{nm}$).  This can be achieved using measurements for which accurate determination of laser intensity is not required, and most of the required measurements are already in the literature.  Measurement of a laser-induced ac-Stark shift of the clock transition would then provide an \emph{in situ} measurement of the laser's intensity to the same $0.5\%$ level of inaccuracy, which is not easily achieved by other means.  This would allow accurate polarisability measurements for clock transitions in other ions, through comparison with $^{138}$Ba$^+$.  The approach would be equally applicable to Sr$^+$ and Ca$^+$, with the latter being immediately applicable to Al$^+$/Ca$^+$ quantum logic clocks.
\end{abstract}
\maketitle
The dynamic differential scalar polarisability $\Delta \alpha_0(\omega)$ of a clock transition is an important quantity to determine, with the dc value $\Delta \alpha_0(0)$ quantifying the blackbody radiation (BBR) shift.  Uncertainty in $\Delta \alpha_0(0)$ is a significant component in the error budget of the Al$^+$ clock \cite{brewer2019quantum}, and likely a limiting factor in Yb$^+$ \cite{huntemann2016single} as well as upcoming multi-ion implementations based on In$^+$ \cite{keller2019controlling} or Lu$^+$ \cite{tan2019suppressing, arnold2018blackbody}.  Although accurate assessments of $\Delta \alpha_0(0)$ have been achieved for ions \cite{dube2014high, huang40}, these determinations relied on having $\Delta \alpha_0(0) < 0$.  Other ion-based clocks have needed to rely on some form of extrapolation from measurements in the near-infrared (NIR) \cite{huntemann2016single, rosenband2006blackbody} and/or by measurement at infrared (IR) wavelengths near to the center of the blackbody spectrum \cite{arnold2018blackbody, arnold2019dynamic,baynham2018measurement}.

The accuracy of polarizability measurements at NIR or IR wavelengths is limited to the accuracy by which the intensity of the laser at the ion can be determined.   This is primarily limited by detector calibration, which is limited to the 1\% level and not always readily available.  Even if the detector is accurately calibrated, the mode of the laser field at the ion must be equally well-calibrated, which is complicated by beam aberration and etaloning effects \cite{arnold2019dynamic}.  Consequently, the ability to accurately determine $\Delta \alpha_0(\omega)$ through intensity-independent measurements is an attractive alternative, which would also allow subsequent \emph{in situ} calibration of laser intensities for measurements against other ions.  Here it is shown that the simple atomic structure of alkaline-earth ions allows such an approach.  Moreover, most of the required measurements have already been reported in the literature.  Although the discussion is focussed on $^{138}$Ba$^+$, the idea is equally applicable to $^{88}$Sr$^+$ and $^{40}$Ca$^+$.

For the $S_{1/2}$ to $D_{5/2}$ transition in $^{138}$Ba$^{+}$, $\Delta \alpha_0(\omega)$ is predominately determined by three transitions at $614, 493$, and $455\,\mathrm{nm}$ with all other contributing transitions having wavelengths below $240\,\mathrm{nm}$. For wavelengths above $700\,\mathrm{nm}$, the ultraviolet (uv) contributions can be well represented by a weak quadratic correction.  Consequently, accurate determination of the matrix elements associated with the three dominant poles, together with a characterisation of an overall dc offset, should provide a reasonably accurate representation of $\Delta \alpha_0(\omega)$ over a wide frequency range.  

In table~\ref{alpha}, contributions to $\Delta \alpha_0(\omega)$ are tabulated using matrix elements calculated by a linearized coupled-cluster method described in Ref.~\cite{safronova2008all} with the exception of the $6s-6p$ transitions which are taken from experiment \cite{woods2010dipole}. The contributions labeled other are obtained using the approach from Refs.~\cite{safronova2009development,porsev1999electric}, after subtracting off leading contributions.  Not given in the table are the core polarizability terms as these are the same for the two states and cancel for a differential polarizability.  However, valence-core correction terms, $\alpha_{vc}$, which compensate for Pauli-principle-violating excitations from the core to the valence shell \cite{safronova1999relativistic}, are included.  Theoretical calculations of matrix elements, polarizabilities and their accuracy are discussed in Appendix \ref{Theory}.
\begin{table}
\caption{Contributions to the differential scalar polarizabilities of the Ba$^+$ clock transition. Dipole matrix elements and polarizability contributions are given in atomic units.}
\begin{ruledtabular}
\begin{tabular}{llcrd}
\multicolumn{1}{l}{State}
&\multicolumn{1}{l}{Contribution} & \multicolumn{1}{c}{$\lambda$ (nm)}
& \multicolumn{1}{c}{$D$} &
\multicolumn{1}{c}{$\alpha$}\\
\hline \\ [-0.3pc]
$6s\;^2\!S_{1/2}$ &$6p\;^2\!P_{1/2}$& 493.5  & 3.3251  &   39.92    \\
                &$7p\;^2\!P_{1/2}$& 202.5  & 0.061  &   0.006    \\
                &$8p\;^2\!P_{1/2}$& 163.0  & 0.087  &   0.009    \\
                &$6p\;^2\!P_{3/2}$& 455.5  & 4.7017  &   73.67     \\
                &$7p\;^2\!P_{3/2}$& 200.0  & 0.087  &   0.011    \\
                &$8p\;^2\!P_{3/2}$& 162.2  & 0.033  &   0.003    \\
                & Other & $<147.8$ & & 0.035\\
                &$\alpha_\mathrm{vc}$      &            &          &  -0.51    \\
                & Total                   &            &          &  113.14    \\[0.5pc]

$5d\;^2\!D_{5/2}$ &$6p\;^2\!P_{3/2}$& 614.3  & 4.103  &   25.22       \\
                &$7p\;^2\!P_{3/2}$& 225.5  & 0.451  &   0.11      \\
                &$8p\;^2\!P_{3/2}$& 178.7  & 0.223  &   0.02      \\
                & Other & $<161.3$ & & 0.04\\
                & Total ($J=3/2$)                          &       &        &  25.39   \\[0.3pc]
                
                &$4f\;^2\!F_{5/2}$& 234.8  & 0.998  &   0.57      \\
	        &$5f\;^2\!F_{5/2}$& 193.4  & 0.220  &   0.02      \\
                &$6f\;^2\!F_{5/2}$& 169.7  & 0.239  &   0.02     \\
                &$7f\;^2\!F_{5/2}$& 157.4  & 0.116  &   0.005    \\
                & Other & $<150.5$ & & 0.10\\
                & Total ($J=5/2$)                          &       &        &  0.725    \\[0.3pc]
                
                &$4f\;^2\!F_{7/2}$& 233.6  & 4.475  &   11.41      \\
	        &$5f\;^2\!F_{7/2}$& 192.5  & 1.089  &   0.56      \\
                &$6f\;^2\!F_{7/2}$& 169.4  & 0.971  &   0.39      \\
                &$7f\;^2\!F_{7/2}$& 157.3  & 0.932  &   0.33      \\
                & Other & $<150.4$ & & 2.02\\
                & Total ($J=7/2$)&       &        &  14.71    \\[0.3pc]
                &$\alpha_\mathrm{vc}$       &       &        &  -0.82     \\
                & Total                           &       &         &  40.00    \\[0.5pc]
\end{tabular}
\end{ruledtabular}
\label{alpha}
\end{table}

\begin{figure}[h]
\begin{center}
  \includegraphics[width=2.8in]{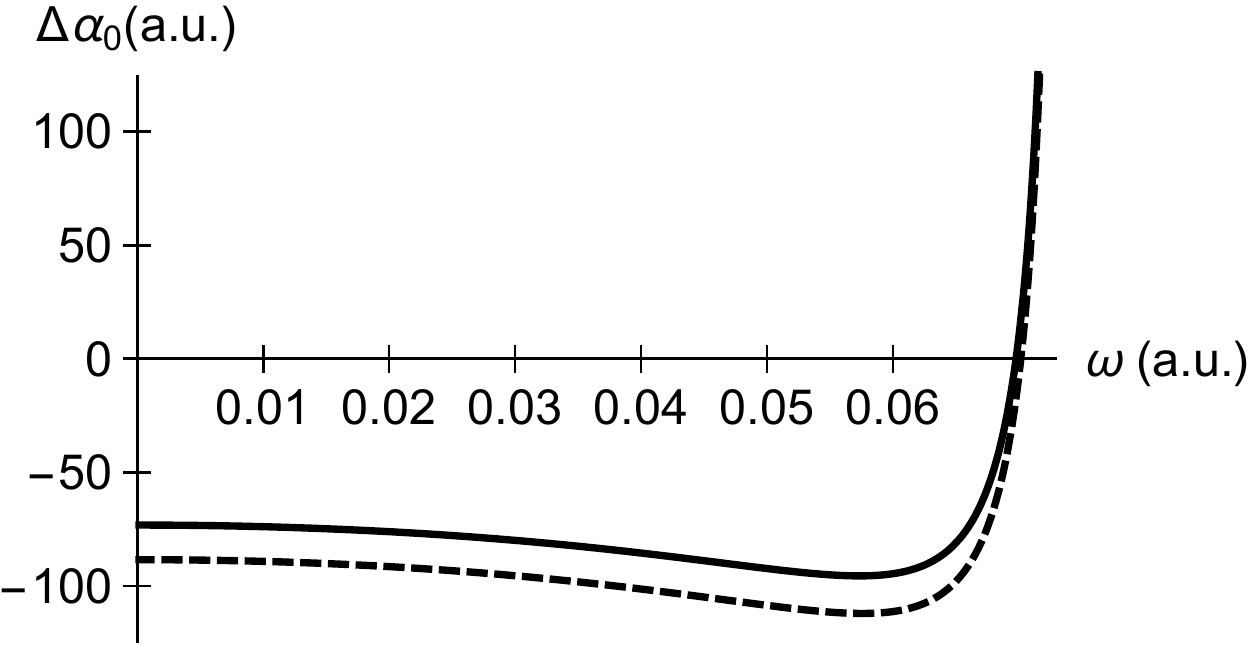}
  \caption{Plot of the differential scalar polarisability, $\Delta\alpha_0(\omega)$.  Solid curve is calculated using matrix elements given in table~\ref{alpha}.  Dashed curve is the contribution from the transitions at 455, 493, and 614\,nm. Both axes are given in atomic units.  Accurate location of the zero crossing at $\omega\approx 0.07\,\mathrm{a.u.}$ ($\lambda\approx653\,\mathrm{nm}$) can facilitate determination of the offset.}
  \label{pol}
\end{center}
\end{figure}

The actual values of the matrix elements and correction terms are not crucial.  More important is that the dominant contributions are determined by the three poles at $614, 493$, and $455\,\mathrm{nm}$ and the rest can be approximated by a weak quadratic form.  This is illustrated in Fig.~\ref{pol}, which shows the polarisability curve calculated from the values given in table~\ref{alpha} and, for comparison, the contribution from the three dominant poles only.  For this purpose the `other' contributions for each intermediate $J$ have been treated as a single pole with the largest possible wavelength of the contributing terms used as the pole position.
As is evident from the figure, in the region $\omega\lesssim0.065\,\mathrm{a.u.}$ ($\lambda \gtrsim 700\,\mathrm{nm}$), the remaining contributions provide an essentially constant offset.

For $\omega\lesssim0.065\,\mathrm{a.u.}$, contributions from the uv transitions and $\alpha_{vc}$ terms can be well approximated by an even-order quadratic polynomial, as can the positive sum of such terms.  From the parameterisation $c_0+c_0\left(\omega/\omega_0\right)^2$, it is readily seen that the single pole
\begin{equation}
\frac{c_0}{1-(\omega/\omega_0)^2},
\end{equation}
has the same quadratic expansion as a sum of poles.  Thus, a single pole can well approximate the uv contributions and $\alpha_{vc}$ terms up to second order.  Additionally, the single pole will partially capture contributions from higher order terms.  Provided there is no significant cancellation of poles, this argument  also holds for a differential polarisability. This is the case for Ba$^+$, as the uv terms are dominated by the $D_{5/2}$ to $4f\;^2\!F_{7/2}$ transition and there is only a few percent contribution from transitions connected to the ground state \cite{iskrenova2008theoretical}.  It therefore follows that, in the region $\omega\lesssim0.065\,\mathrm{a.u.}$, $\Delta \alpha_0(\omega)$ can be well approximated by a sum of four poles:
\begin{align}
\Delta\alpha_0(\omega)&=\Delta\alpha_0^\mathrm{vis}(\omega)+\Delta\alpha_0^\mathrm{uv}(\omega)\nonumber\\
&\approx\Delta\alpha_0^\mathrm{vis}(\omega)+\frac{c_0}{1-(\omega/\omega_0)^2},
\label{approxEq}
\end{align} 
where $\Delta\alpha_0^\mathrm{vis}(\omega)$ gives the contributions from the 455, 493, and 614\,nm transitions, and $\Delta\alpha_0^\mathrm{uv}(\omega)$ the rest.

Mathematically, the approximation given by Eq.~\ref{approxEq} can be exceptionally good.  Taking $\Delta \alpha_0(\omega)$ calculated using all contributions given in table~\ref{alpha} as a representative example,  $c_0$  and $\omega_0$ can be chosen such that the approximation matches the zero crossing of $\Delta \alpha_0(\omega)$ and minimizes the discrepancy over the range $\omega_0<0.065\,\mathrm{a.u.}$  This gives $c_0=15.23\mathrm{a.u.}$ and $\omega_0=0.20479\,\mathrm{a.u.}$ ($\lambda_0=222.49\,\mathrm{nm}$), with a maximum fractional discrepancy of $\sim 2\times 10^{-5}$ over the frequency range of interest.  The agreement only relies on the validity of the single pole approximation to the uv and $\alpha_{vc}$ terms, which is not dependent on exact values of matrix elements.  The practical limitation is set by how well the approximation can be realized.

To experimentally characterize the approximation, the procedure would be to first fix the three main contributions, by determining directly the matrix elements associated with the transitions at 455, 493, and 614\,nm, and then to locate the zero crossing to determine the offset.  Since the quadratic correction is weak, the quality of the approximation is insensitive to $\omega_0$, so it can be fixed to a value determined by theory.  With $\omega_0$ fixed, $c_0$ would then be chosen so that the zero crossing for the approximation matches the measured position of the zero crossing at $\omega\approx 0.07\,\mathrm{a.u.}$ ($\lambda\approx653\,\mathrm{nm}$). It then remains to determine how good the approximation is, taking into account reasonable experimental measurements and theoretical estimates of $\omega_0$.
 
High accuracy determination of individual matrix elements has been achieved in a number of different ways. Precision measurement of excited state lifetimes \cite{olmschenk2006precision,andra1973new,jin1993precision} and branching fractions \cite{ramm2013precision,dutta2016exacting,de2015precision} give matrix elements with inaccuracies $\lesssim 1\%$. For $^{40}$Ca$^+$, comparison of off-resonant scattering rates and Stark shifts enabled the determination of matrix elements with inaccuracies at the $0.1\%$ level \cite{hettrich2015measurement}.  For Ba$^+$, resonant excitation Stark ionisation spectroscopy has been used to determine matrix elements for the $493$ and $455\,\mathrm{nm}$ transitions with reported inaccuracies of $0.05\%$ \cite{woods2010dipole}.  The latter measurements, combined with branching fractions given in \cite{dutta2016exacting}, would determine the matrix element $\langle P_{3/2}\|r\|D_{5/2}\rangle$ to an inaccuracy of $\sim0.3\%$, although from \cite{arnold2019measurements} an independent assessment would be of interest, in addition to an improved accuracy.

Improved accuracy of $\langle P_{3/2}\|r\|D_{5/2}\rangle$ with an independent assessment should be readily achievable.  Optical pumping into $D_{3/2}$ followed by depumping with $585\,\mathrm{nm}$ light, which couples $D_{3/2}$ to $P_{3/2}$, would optically pump the atom into $S_{1/2}$ and $D_{5/2}$ with probability $p\sim0.66$ and $1-p$, respectively.  Measurement of $p$ would then provide the desired matrix element via the relation
\begin{equation}
\frac{\langle P_{3/2}\|r\|D_{5/2}\rangle}{\langle P_{3/2}\|r\|S_{1/2}\rangle}=\left(\frac{\omega_{455}}{\omega_{614}}\right)^{3/2}\sqrt{\frac{1-p}{p}}.
\end{equation}
The fractional inaccuracy in the determination of $\langle P_{3/2}\|r\|D_{5/2}\rangle$ due to projection noise in a measurement of $p$ is then $\sim1/\sqrt{N}$, where $N$ is the number of measurements.  This method is insensitive to laser intensities, polarisation, and detunings.  Since $D_{5/2}$ has a lifetime of $\sim 30\,\mathrm{s}$, state detection errors can be negligibly small and accuracy would be ultimately limited by the accuracy of $\langle P_{3/2}\|r\|S_{1/2}\rangle$.  Thus this contribution could also be determined to an inaccuracy at the $0.1\%$ level.

For a given value of $\omega_0$, $c_0$ can be set by determining the zero crossing near $653\,\mathrm{nm}$.  In this region there is a large contribution from the tensor polarisability, but this can be heavily suppressed by appropriate orientation of the magnetic field with respect to the laser polarisation, as done in recent experiments with Lu$^+$ \cite{arnold2018blackbody}, and by averaging over Zeeman pairs, as done with Sr$^+$ \cite{dube2005electric}.  Also, determination of the zero crossing does not require an accurate assessment of laser intensity.  At $\pm500\,\mathrm{GHz}$ from the zero crossing, $\Delta \alpha_0(\omega)\approx \pm 3\,\mathrm{a.u.}$, which should enable a readily measurable Stark shift.  Linear interpolation of the two points would then give an estimate of the zero point.  Provided the intensity was stabilised to a fixed value for both measurements, accuracy of this approach would be limited by the curvature of $\alpha_0(\omega)$ within this region, which would bias the result by an estimated $\approx -10\,\mathrm{GHz}$.  Based on this, 20\,GHz should be an achievable uncertainty for the zero crossing.

Determination of $\omega_0$ would rely on theoretical calculations.  From Eq.~\ref{approxEq}, the zero crossing $\Delta \alpha_0(\omega')=0$ gives 
\begin{equation}
\label{uvEq}
\Delta\alpha_0^\mathrm{uv}(\omega')=-\Delta\alpha_0^\mathrm{vis}(\omega')\approx\frac{c_0}{1-(\omega'/\omega_0)^2}.
\end{equation} 
Since $\Delta\alpha_0^\mathrm{vis}(\omega')$ can be determined accurately by independent measurements, locating the zero crossing constitutes a constraint on $\Delta\alpha_0^\mathrm{uv}(\omega')$.  A theoretical estimate for $\omega_0$ and its uncertainty can be found by solving Eq.~\ref{uvEq} using calculated values of $c_0=\Delta\alpha_0^\mathrm{uv}(0)$ and $\Delta\alpha_0^\mathrm{uv}(\omega')$ with $c_0$ allowed to vary subject to a constraint on $\Delta\alpha_0^\mathrm{uv}(\omega')$.  This is effectively an extrapolation to dc based on a measurement at $\omega'$.

To illustrate, $\Delta\alpha_0^\mathrm{uv}(\omega)$ is first written in the vector form
\begin{equation}
\Delta\alpha_0^\mathrm{uv}(\omega)=\sum_k\frac{c_k}{1-(\omega/\omega_k)^2}=\mathbf{f}(\omega)\cdot \mathbf{c}
\end{equation}
where the $k^{th}$ component of $\mathbf{f}(\omega)$ is $1/(1-(\omega/\omega_k)^2)$ and the summation is over all contributing uv transitions, with constant terms having $\omega_k\rightarrow \infty$.  The coefficients $\mathbf{c}$ have theoretical estimates $\mathbf{c}_0$, with uncertainties $\delta \mathbf{c}$, and, since transition frequencies are generally well-known, $\mathbf{f}(\omega)$ is practically exact.  To find the allowable variation in $\Delta\alpha_0^\mathrm{uv}(0)$ consistent with $\Delta\alpha_0^\mathrm{uv}(\omega')$, $\mathbf{f}(0)$ is written as a projection onto $\mathbf{f}(\omega')$ and an orthogonal unit vector $\hat{\mathbf{n}}$, i.e. $\mathbf{f}(0)= a_1 \, \mathbf{f}(\omega')+a_2 \, \hat{\mathbf{n}}$, which gives
\begin{equation}
\Delta\alpha_0^\mathrm{uv}(0)=\mathbf{f}(0)\cdot \mathbf{c}= a_1 \Delta\alpha_0^\mathrm{uv}(\omega')+a_2\, \hat{\mathbf{n}}\cdot\mathbf{c}.
\end{equation}
This expression has the same form as that used in the assessment of the BBR shift in the Al$^+$ clock \cite[Eq. S24]{brewer2019quantum}.  Here, by construction, the two terms are independent as required for uncertainties to be added in quadrature.  Variations in $\Delta\alpha_0^\mathrm{uv}(0)$ due to those in $\Delta\alpha_0^\mathrm{uv}(\omega')$ do not significantly influence an estimation of $\omega_0$.  Hence, we allow $c_0$ to vary by $a_2\, \hat{\mathbf{n}}\cdot\delta\mathbf{c}$ in the application of Eq.~\ref{uvEq} to determine an uncertainty in $\omega_0$.  As this is a theoretical determination, $\Delta\alpha_0^\mathrm{uv}(\omega')$ is fixed to that estimated at the theoretically determined zero crossing.  

Using the above approach and the values in table~\ref{alpha}, we find $\omega_0=0.2049(42)\,\mathrm{a.u.}$ (222(5)\,nm), where we have used a 4\% uncertainty in the two $4f$ contributions and a 100\% uncertainty in all others.  As before, the contributions labelled `other' have been treated as single poles.  Consequently, the errors in the contributions from these terms are assumed correlated.  Correlation is also assumed for the errors in the $nF_{5/2}$ and $nF_{7/2}$ contributions as these are expected to be related.  The assumed frequency dependence of the `other' terms does not significantly affect the uncertainty derived in $\omega_0$.  Therefore, $\pm5\,\mathrm{nm}$ is taken as a reasonable uncertainty for the pole placement. 

We stress that the value of $c_0$ used in Eq.~\ref{approxEq} would be chosen so that, for the given estimated $\omega_0$, the expression would give the same measured zero crossing.  Location of the zero crossing would need to be consistent with that estimated from theory, which we calculate to be $653.0(1.3)\,\mathrm{nm}$.  If this were not the case, there would be no justification for asserting the validity of the estimate of $\omega_0$.  However, such an inconsistency would be rather surprising, given the agreement between theory and experiment for matrix elements \cite{woods2010dipole,arnold2019measurements}, branching ratios \cite{dutta2016exacting,de2015precision,arnold2019measurements}, and even $\Delta\alpha_0(0)$ \cite{yu1994stark}, although the latter has a large uncertainty.

\begin{figure}[h]
\begin{center}
  \includegraphics[width=3in]{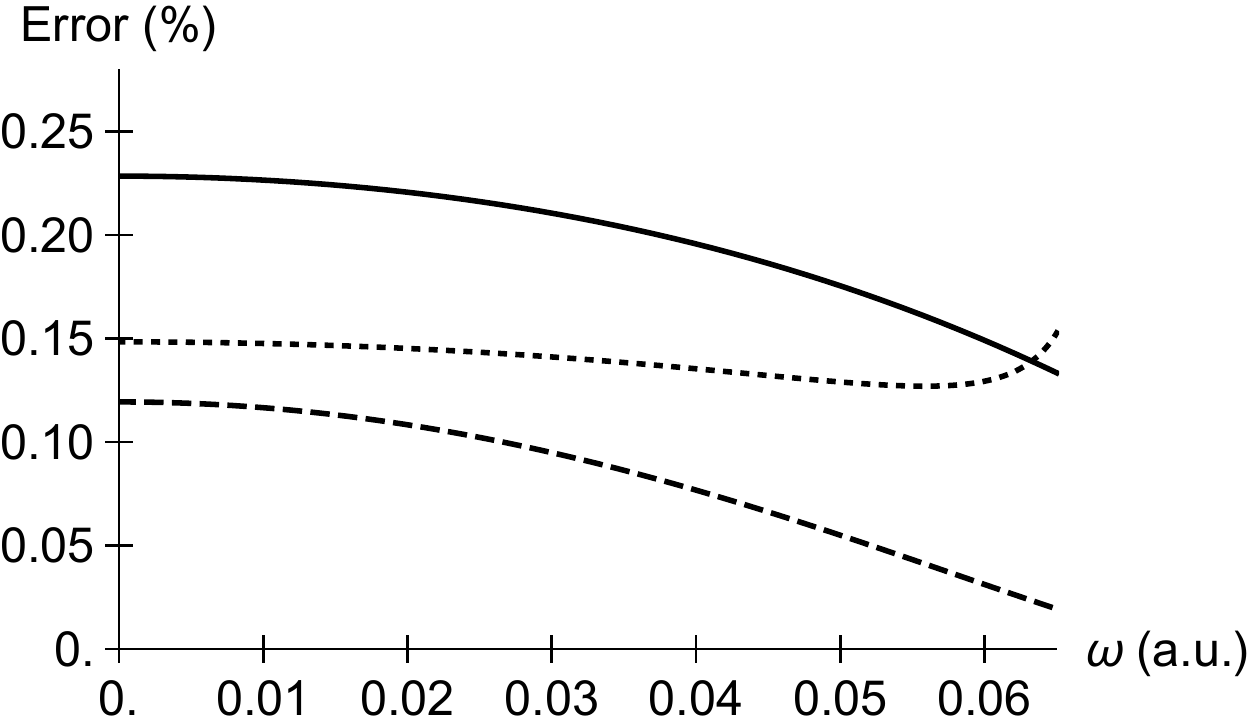}
  \caption{Plot of the percentage error contributions between a full calculation of $\Delta\alpha_0(\omega)$ and the approximate expression given in Eq.~\ref{approxEq} as as a function of the angular frequency $\omega$ given in atomic units. The solid curve is the error contribution if the pole strength for the 614\,nm transition is decreased by $0.1\%$.  The dashed curve is the error contribution if the uv pole is shifted by 5\,nm to 217.5\,nm.  The dotted curve is the error contribution if the zero-crossing is underestimated by $20\,\mathrm{GHz}$.}
  \label{polE}
\end{center}
\end{figure}

To illustrate the sensitivity to the various error contributions, the fractional difference between $\Delta \alpha_0(\omega)$ calculated using all contributions and the approximation given in Eq.~\ref{approxEq} for various errors is plotted  in Fig.~\ref{polE}.  The solid curve is the error introduced with a fractional decrease of $10^{-3}$ in the 614-nm contribution, the dashed curve is the error contribution if the uv pole is shifted to 217.5\,nm, and the dotted curve is the error contribution if the zero crossing is underestimated by $20\,\mathrm{GHz}$.  Each curve scales almost linearly with the stated error, such that the result of a change in sign of an error can be approximated by a reflection of the associated curve about the horizontal axis.  Errors arising from the 455 and 493 transitions have been omitted as they are smaller by a factor of $\sim3$ than that from the 614-nm transition.  This is due to the relative position of the transition with respect to the zero crossing. Adding errors in quadrature, including those from the 455 and 493 poles, gives a maximum error of $0.32\%$ over the entire region from dc to $\omega=0.065\,\mathrm{a. u.}$ ($\sim 700\,\mathrm{nm}$).  

As already noted, the reduced matrix elements $\langle P_{3/2}\|r\|S_{1/2}\rangle$ and $\langle P_{1/2}\|r\|S_{1/2}\rangle$ have already been reported in the literature with inaccuracies of $\sim0.05\%$ \cite{woods2010dipole}.  Hence, all that remains is an improved measurement of the branching fraction $p$ and location of the zero crossing near $653\,\mathrm{nm}$.  In addition, $\Delta\alpha_0(0)<0$, which should allow a high accuracy measurement of $\Delta\alpha_0(0)$ as done with Sr$^+$ and Ca$^+$ \cite{dube2014high,huang40}.  This would provide a rigorous consistency check among multiple precision measurements and an experimental assessment of $\omega_0$.

In summary, we have shown that the dynamic differential scalar polarisability, $\Delta\alpha_0(\omega)$, of the $S_{1/2}-D_{5/2}$ transition in $^{138}$Ba$^+$ can be determined to an inaccuracy below $0.5\%$ across a wide wavelength range ($\lambda>700\,\mathrm{nm}$).  Moreover, the determination can be obtained using measurements that do not require accurate determination of laser intensities and some of the required measurements have already been reported in the literature.  Although the method relies on a theoretical estimate of an effective pole position $\omega_0$, the resulting approximation to $\Delta\alpha_0(\omega)$ is relatively insensitive to this value such that this is unlikely to be a significant limitation.

The methodology proposed here would also be applicable to Sr$^+$ and Ca$^+$.  For these cases, uv transitions are deeper in the uv making the approximation less sensitive to the choice of $\omega_0$.  In the case of Ca$^+$, an accurate measurement of the $\langle P_{1/2}\|r\|S_{1/2}\rangle$ matrix element has been reported \cite{hettrich2015measurement} and $\langle P_{3/2}\|r\|S_{1/2}\rangle$ can be well-approximated by $\langle P_{1/2}\|r\|S_{1/2}\rangle\sqrt{2}$ \cite{safronova2011blackbody}.  Together with the branching fractions reported in \cite{gerritsma2008precision}, and the recent high accuracy determination of $\Delta\alpha_0(0)$ \cite{huang40}, a calibration of the polarizability curve to $\lesssim1\%$ could be done.  Determination of the zero crossing, which we estimate to be at 297.5(2)\,THz, would provide a consistency check of the methodology.   

The case of Ca$^+$ is of particular relevance to the Al$^+$ clock, for which the uncertainty in $\Delta\alpha_0(0)$ is now a significant contribution to the error budget \cite{brewer2019quantum}.  Measurement of this quantity has only been carried out twice and both rely on extrapolation from a single measurement point \cite{rosenband2006blackbody,brewer2019quantum}.  It is therefore desirable to provide an independent assessment.  Clock implementations utilizing Ca$^+$ as the logic ion would allow accurate calibration of a laser intensity at multiple wavelengths and improved measurements of $\Delta\alpha_0(\omega)$ for Al$^+$.  For wavelengths above 780\,nm, the differential scalar polarizability of the Al$^+$ clock transition is well approximated by a quadratic form and even two measurements of $\Delta\alpha_0(\omega)$ in the NIR would allow a more accurate extrapolation to dc.

Measurements proposed in this work will also provide benchmarks for matrix elements involving 4f states, as needed in calculations for highly-charged ions. In addition, they will provide a precision test of methods to compute polarizability contributions from highly-excited states, which will be useful in establishing theoretical uncertainties of predicted polarizabilities in other systems.
\appendix 
\section{Matrix element calculations}
\label{Theory}
\begin{table*}[!htbp]
\caption{\label{tab1} Absolute values of the reduced matrix elements contributing to the $5d_{5/2}$ polarizability calculated in different approximations (in a.u.). DHF - Dirac Hartree-Fock lowest order, RPA - random phase approximation, RPA+$\Sigma_1^{\text{(2,all)}}$ include correlation potential in second and all-order approximations, respectively. The all-order single-double (SD) and single-double + partial triple (SDpT) results are listed in SD and
 SDpT columns, corresponding scaled vales are listed in the SD$_{\text{sc}}$ and SDpT$_{\text{sc}}$ columns. Uncertainties are given in parentheses. *See text for a discission of uncertainties.}
\begin{ruledtabular}
\begin{tabular}{lccccccccc}
\multicolumn{1}{c}{Transition} &
\multicolumn{1}{c}{DHF} &
\multicolumn{1}{c}{RPA} &
\multicolumn{1}{c}{RPA+$\Sigma_1^{\text{(2)}}$} &
\multicolumn{1}{c}{RPA+$\Sigma_1^{\text{(all)}}$} &
\multicolumn{1}{c}{SD} &
\multicolumn{1}{c}{SDpT} &
\multicolumn{1}{c}{SD$_{\text{sc}}$} &
\multicolumn{1}{c}{SDpT$_{\text{sc}}$} &
\multicolumn{1}{c}{Final}  \\
\hline
$5d_{5/2} - 6p_{3/2}$&	4.993	&4.592&	4.015&	4.090&	4.103	&4.163&	4.137&	4.122 &  4.103(50)\\		
$5d_{5/2} - 7p_{3/2}$&	0.546	&0.368&	0.424&	0.422&	0.451	&0.450&	0.446&	0.457 &  0.451(9)\\		
$5d_{5/2} - 8p_{3/2}$&	0.299	&0.187&	0.207&	0.205&	0.223	&0.224&	0.221&	0.225 &  0.223(4)\\	[0.5pc]																	  	
$5d_{5/2} - 4f_{5/2}$&	1.145	&1.040&	0.955&	0.986&  0.998	&1.012&	1.011&	1.009&  0.998(20)\\		
$5d_{5/2} - 5f_{5/2}$&	0.629	&0.537&	0.159&	0.102&	0.016	&0.210&	0.027&	0.220 &  0.220*	\\	
$5d_{5/2} - 6f_{5/2}$&	0.406	&0.330&	0.239&	0.262&	0.236	&0.018&	0.239&	0.024 &  0.239*\\
$5d_{5/2} - 7f_{5/2}$&	0.286	&0.223&	0.195&	0.221&	0.113	&0.116&	0.116&	0.108 &   0.12(6)	\\	[0.5pc]						       		 	
$5d_{5/2} - 4f_{7/2}$&	5.128	&4.655&	4.335&	4.464&	4.475	&4.540&	4.521&	4.523 &       4.475(90)\\		
$5d_{5/2} - 5f_{7/2}$&	2.812	&2.402&	0.520&	0.236&	0.130	&1.049&	0.085&	1.089 &       1.089*\\
$5d_{5/2} - 6f_{7/2}$&	1.815	&1.475&	0.999&	1.086&	0.961	&0.170&	0.971&	0.186 &     0.971*	\\
$5d_{5/2} - 7f_{7/2}$&	1.278	&0.996&	0.838&	0.952&	0.922	&0.429&	0.932&	0.388 &   0.93(54)	
\end{tabular}
\end{ruledtabular}
\end{table*}

\begin{table} [h]
\caption{\label{tab2}Contributions to the static scalar $6s$ polarizability $\alpha_0(0)$ and dynamic polarizability
at $\lambda=653.0$~nm. The absolute values of the $6s-np$ reduced matrix elements (in a.u.)
are also listed in the column labelled ME. Uncertainties are given in parentheses. }
\begin{ruledtabular}
\begin{tabular}{lccc}
\multicolumn{1}{c}{Contribution} &
\multicolumn{1}{c}{ME} &
\multicolumn{1}{c}{$\alpha_0(0)$} &
\multicolumn{1}{c}{$\alpha_0(\omega)$} \\
\hline
$6p_{1/2}$      & 3.3251(21)$^{(a)}$ &   39.921(48)&    93.11(11)    \\
$7p_{1/2}$      &    0.061   &    0.006    &   0.006        \\
$8p_{1/2}$      &    0.087   &    0.009    &   0.010        \\
$(n>8)p_{1/2}$  &            &   0.030(20) &   0.030(20)    \\   [0.5pc]	
$6p_{3/2}$     &  4.7017(27)$^{(a)}$ &   73.670(88)  &  143.51(17)   \\
$7p_{3/2}$     &  0.087      &   0.011       &   0.012      \\
$8p_{3/2}$     &   0.033     &   0.003       &   0.001      \\
$(n>8)p_{3/2}$ &   0.0057    &   0.005(20)   &   0.005(20)  \\   [0.5pc]	
$\alpha_{vc}$   &             &  -0.51(13)  &   -0.51(13)    \\
Sum             &             &  113.14(17) &   236.17(25)  \\
Core            &             &  10.6(5)    &   10.6(5)     \\
Final           &             &  123.7(5)   &    246.8(6)   \\
\end{tabular}
\end{ruledtabular}
\begin{flushleft}
$^{(a)}$Ref.~\cite{woods2010dipole}.
\end{flushleft}
\end{table}

\begin{table} [htbp]
\caption{\label{tab3} The tail contribution to the static scalar polarizability
of the $5d_{5/2}$ state calculated in different approximations (in a.u.). The contributions the
  from $(n>8)p_{3/2}$  and $(n>7)f_{j}$ higher states are included. The same designations are used as in Table~\ref{tab1}.  }
\begin{ruledtabular}
\begin{tabular}{lc}
\multicolumn{1}{c}{Approximation} &
\multicolumn{1}{c}{Tail} \\
\hline
DHF			&2.498\\
RPA			&1.817\\
RPA+$\Sigma_1^{\text{(2)}}$	&		2.249\\
RPA+$\Sigma_1^{\text{all}}$&			2.156\\
Final&2.16(34)
\end{tabular}
\end{ruledtabular}
\end{table}

\begin{table} [htbp]
\caption{\label{tab4} Contributions to the static scalar $5d_{5/2}$ polarizability $\alpha_0(0)$ and dynamic polarizability
at $\lambda=653.0$~nm.  Uncertainties are given in parentheses. }
\begin{ruledtabular}
\begin{tabular}{lcc}
\multicolumn{1}{c}{Contribution} &
\multicolumn{1}{c}{$\alpha_0(0)$} &
\multicolumn{1}{c}{$\alpha_0(\omega)$} \\
\hline $6p_{3/2}$       &     25.22(61)  &       219.5(5.3)   \\
$7p_{3/2}$        &    0.112(5)    &     0.127(5)  \\
$8p_{3/2}$        &    0.022(1)    &     0.023(1)  \\
$(n>8)p_{3/2}$    &      0.037     &       0.038   \\   [0.5pc]
$4f_{5/2}$       &     0.570(23)  &    0.655(26)   \\
$5f_{5/2}$       &     0.023(23)  &    0.025(25)   \\
$6f_{5/2}$       &     0.024(24)  &    0.025(25)   \\
$7f_{5/2}$       &     0.005(5)   &    0.006(6)    \\
$(n>7)f_{5/2}$    &       0.103    &     0.106      \\ [0.5pc]
$4f_{7/2}$        &     11.41(46)  &     13.08(52) \\
$5f_{7/2}$        &     0.56(56)   &    0.61(61)   \\
$6f_{7/2}$        &     0.39(39)   &    0.42(42)   \\
$7f_{7/2}$        &     0.33(33)   &    0.35(35)   \\
$(n>7)f_{7/2}$    &     2.02       &  2.09         \\       [0.5pc]
Total tail        &    2.16(34)    &  2.23(34)     \\
$\alpha_{vc}$     &    -0.82(3)    &     -0.82(3)  \\
   Total            &     40.0(1.1)  &     236.2(5.4) \\
Core             &    10.6(5)     &   10.6(5)       \\
Final            &    50.6(1.2)   &   246.8(5.4)   \\
\end{tabular}
\end{ruledtabular}
\begin{flushleft}
\end{flushleft}
\end{table}

The valence parts of the scalar, $\alpha_0$, and tensor, $\alpha_2$, polarizabilities of Ba$^+$ levels may be calculated using the sum-over-states
expressions \cite{MitSafCla10}:
\begin{eqnarray}
    \alpha_{0}^v(\omega)&=&\frac{2}{3(2j_v+1)}\sum_k\frac{{\left\langle k\left\|D\right\|v\right\rangle}^2 \Delta E}{\Delta E^2-\omega^2}, \rm{~and} \label{eq-1} \nonumber \\
    \alpha_{2}^v(\omega)&=&-4C\sum_k(-1)^{j_v+j_k+1}
            \left\{
                    \begin{array}{ccc}
                    j_v & 1 & j_k \\
                    1 & j_v & 2 \\
                    \end{array}
            \right\} \nonumber \\
      & &\times \frac{{\left\langle
            k\left\|D\right\|v\right\rangle}^2\Delta E}{
            \Delta E^2-\omega^2} \label{eq-pol},
\end{eqnarray}
             where $C$ is given by
\begin{equation}
            C =
                \left(\frac{5j_v(2j_v-1)}{6(j_v+1)(2j_v+1)(2j_v+3)}\right)^{1/2}. \nonumber
\end{equation}
Here, $\delta_E=E_k-E_v$, $\left\langle i\left\|D\right\|j\right\rangle$ are reduced electric-dipole matrix elements and the sum over intermediate $k$ states includes contributions from all transitions allowed by the electric-dipole selection rules. We use a finite B-spline basis set which make this sum finite. The first few terms give dominant contributions and respective matrix elements have to be calculated with the highest possible accuracy. We use a linearised coupled-cluster (LCC) method \cite{safronova2008all} that includes dominant classes of correlation corrections to all orders of perturbations theory. This method was used for the prediction of the Ca$^+$ \cite{Ca} and Sr$^+$ \cite{Sr} differential clock state scalar polarizabilities and subsequent measurements confirmed the accuracy of this approach.

Four different LCC calculations were carried out: two \textit{ab initio} calculations that include single-double excitations (SD) and additional partial triple contributions (SDpT), and two other calculations, labeled SD$_{\text{sc}}$ and SDpT$_{\text{sc}}$, where higher excitations are estimated using a scaling procedure. Details of the method and a description of the scaling procedures are given in \cite{safronova2008all}.  The all-order results are given in Table~\ref{tab1}. We also list lowest order Dirac Hartree-Fock (DHF) and random phase approximation (RPA) values to demonstrate the size of the correlations corrections. In addition, the matrix elements that include RPA and corrections to the one-body part of the Hamiltonian ($\Sigma_1$) are included. Two ($\Sigma_1$)
 calculations were carried out; one to second order of perturbation theory, and the other to all orders. These calculations follow the methods described in
 \cite{all}, with the valence-valence part of the calculations omitted, as Ba$^+$ has a single valence electron. We use these methods to evaluate polarizability contributions from the higher states and it is important to compared these results to the final LCC values. The uncertainties of the $5d_{5/2} - np_{3/2}$ and $5d_{5/2} - 4f_{j}$ matrix elements are determined as the maximum difference of the final and three other LCC values.

Correlations corrections are very large for the $nf$ Ba$^+$ states, which causes convergence issues in the LCC calculations that cannot be fixed with usual stabiliser methods \cite{stab}. We use additional fitting for the $nf$ states to resolve this issue ensuring correct energies after the termination of the LCC calculations. We still find very large differences between the SD and SDpT $5d_{5/2}-5f_j$ and $5d_{5/2}-6f_j$ values.  As a result, we assign a 100\% uncertainty to the corresponding $5d_{5/2}$ polarizability contributions based on the spread of LCC matrix element values.

The contributions to the $6s$ static and dynamic polarizabilities at $\lambda=653$~nm are given in Table~\ref{tab2}.
Experimental values from~\cite{woods2010dipole} obtained using  the resonant excitation Stark ionization spectroscopy technique are used for the $6s-6p$ matrix elements. Experimental energies are used in the calculation of main contributions for all polarizability calculations. The contribution of states with $n>8$ is very small and is calculated in the RPA. A maximum difference of the DHF and RPA tail values for the $np_{1/2}$ and $np_{3/2}$ cases is taken to be the tail uncertainty. The ionic core polarizabilty and small correction accounting for the occupied valence orbital ($\alpha_{vc}$) are also calculated in the RPA.

Because of significant contributions from the higher $nf_{7/2}$  states to the $5d_{5/2}$ polarizability, we use a more accurate method to evaluate the tail for the $5d_{5/2}$ polarizability. The tail includes the contribution of the $(n>8)p_{3/2}$  and $(n>7)f_{j}$ states. Instead of using the sum-over-states approach we solve the inhomogeneous equation of perturbation theory in the valence space, which is approximated as
\begin{equation}
(E_v - H_{\textrm{eff}})|\Psi(v,M^{\prime})\rangle = D_{\mathrm{eff},q} |\Psi_0(v,J,M)\rangle
\label{eq1}
\end{equation}
for a state $v$ with the total angular momentum $J$ and projection $M$ \cite{kozlov99a} and then use resulting wave functions for the polarizability calculations. The $H_{\textrm{eff}}$ term includes either second-order ($\Sigma_1^{(2)}$) or the all-order ($\Sigma_1^{(all)}$) corrections as described  in \cite{all}, the effective dipole operator $D_{\textrm{eff}}$ includes random phase approximation (RPA) corrections. Tail results, calculated in various approximations, are listed in Table~\ref{tab3}. We find results to be very stable with the approximation and assign the spread of the values as the uncertainty.

The crossing of the $6s$ and $5d_{5/2}$ static polarizabilities is found to be $653.0(1.3)\,\mathrm{nm}$, where the uncertainty is predominately due to the uncertainty of the $5d_{5/2}$ contributions.  As seen in Table~\ref{tab4}, the uncertainty is almost entirely from the $5d_{5/2}-6 p_{3/2}$ contribution.  Thus we can expect this to be improved once a more accurate determination of the $5d_{5/2}-6 p_{3/2}$ matrix element is made.  For completeness we note that the static tensor polarizability and the dynamic tensor polarizability at $653\,\mathrm{nm}$ are calculated to be $-29.8(7)\,\mathrm{a.u.}$ and  $-225(5)\,\mathrm{a.u.}$, respectively.  

\begin{acknowledgements}
This work is supported by the National Research Foundation, Prime Ministers Office, Singapore and the Ministry of Education, Singapore under the Research Centres of Excellence programme, and by A*STAR SERC 2015 Public Sector Research Funding (PSF) Grant (SERC Project No: 1521200080). This research was performed in part under the sponsorship of the Office of Naval Research, USA, Grant No. N00014-17-1-2252.
\end{acknowledgements}

\bibliography{PolarizabilityBa}
\bibliographystyle{unsrt}
\end{document}